\documentclass{PoS}

\usepackage{graphicx}

\title{Recent Belle results in quarkonium physics}

\ShortTitle{Belle results in quarkonium}

\author{\speaker{Roman Mizuk}\\
        Institute for Theoretical and Experimenal Physics, Moscow\\
        E-mail: \email{mizuk@itep.ru}}

\abstract{ We review selected recent results of Belle in quarkonium
  physics, that include precision measurement of the $\eta_b(1S)$
  parameters, evidence for the $\eta_b(2S)$; evidence for the
  $\psi_2(1D)$; observation of the $\psi(4040)$ and $\psi(4160)$
  transitions to $\jp\eta$ with anomalously high rates; observation of
  the $\Uf$ transitions to $\U(1D)\pp$ and $\U(1S,2S)\eta$.
  The low excitations of charmonium and bottomonium are in agreement
  with the Lattice QCD and effective theories calculations, while high
  excitations show unexpected properties.  }
 
\FullConference{Xth Quark Confinement and the Hadron Spectrum,\\
                October 8-12, 2012\\
                Munich, Germany}

\newcommand{\mev}{\mathrm{MeV}}
\newcommand{\kev}{\mathrm{keV}}

\newcommand{\mevm}{\mathrm{MeV}/c^2}

\newcommand{\ee}{e^+e^-}
\newcommand{\uu}{\mu^+\mu^-}
\newcommand{\pp}{\pi^+\pi^-}
\newcommand{\ga}{\gamma}
\newcommand{\U}{\Upsilon}
\newcommand{\Uf}{\Upsilon(5S)}
\newcommand{\Uo}{\Upsilon(1S)}

\newcommand{\Ut}{\Upsilon(2S)}
\newcommand{\Uth}{\Upsilon(3S)}
\newcommand{\Un}{\Upsilon(nS)}

\newcommand{\et}{\eta_b(1S)}
\newcommand{\ett}{\eta_b(2S)}

\newcommand{\etm}{\eta_b(mS)}
\newcommand{\hb}{h_b(1P)}
\newcommand{\hbp}{h_b(2P)}
\newcommand{\hbn}{h_b(nP)}
\newcommand{\hbm}{h_b(mP)}
\newcommand{\jp}{J/\psi}

\newcommand{\psp}{\psi(2S)}

\newcommand{\br}{\mathcal{B}}

\newcommand{\etal}{\em et al.}

\newcommand{\zbo}{Z_b(10610)}
\newcommand{\zbt}{Z_b(10650)}

\begin{document}

B-factories have made a significant contribution to the quarkonium
physics, in particular, they observed several missing low excitations
and many highly excited states with unexpected properties.
In this review of recent Belle results we consider charmonium and
bottomonium states in parallel to stress similarity between the two
sectors. We start from low excitations and then move beyond the open
flavor threshold.

\section{Heavy quarkonia below open flavor thresholds}

\subsection{$\et$ and $\ett$}

Spin-singlet states provide information on spin-spin interaction
between quark and antiquark. Observed in 2008, $\et$ remained the only
known bottomonium spin-singlet state~\cite{et_bbr_cleo}. Recently
Belle observed the $\hb$ and $\hbp$ states using transitions
$\Uf\to\hbn\pp$ that were reconstructed
inclusively~\cite{hb_belle}. The \mbox{$P$-wave} hyperfine splittings
$\Delta M_{HF}(nP)=\sum\limits_{J=0}^2
\frac{2J+1}{9}m_{\chi_{bJ}(nP)}-m_{\hbn}$ are measured to be
$(+0.8\pm1.1)\,\mevm$ for $n=1$ and $(+0.5\pm1.2)\,\mevm$ for
$n=2$~\cite{et_belle}; consistent with zero values are in agreement
with expectations~\cite{hf_pwave_theory,qwg_reports}.

The $\hbn$ production in the dipion transitions from the $\Uf$ is
found to be unsuppressed relative to the $\Un$ production despite
involved spin-flip of heavy quark~\cite{hb_belle}. Further studies of
these transitions resulted in the observation of charged
bottomonium-like states $\zbo$ and
$\zbt$~\cite{zb_belle,zb_belle_angular,zb_belle_bb,zb_belle_neutral},
that are discussed in a separate talk at this conference.

\begin{figure}[tbhp]
\includegraphics[width=0.33\linewidth]{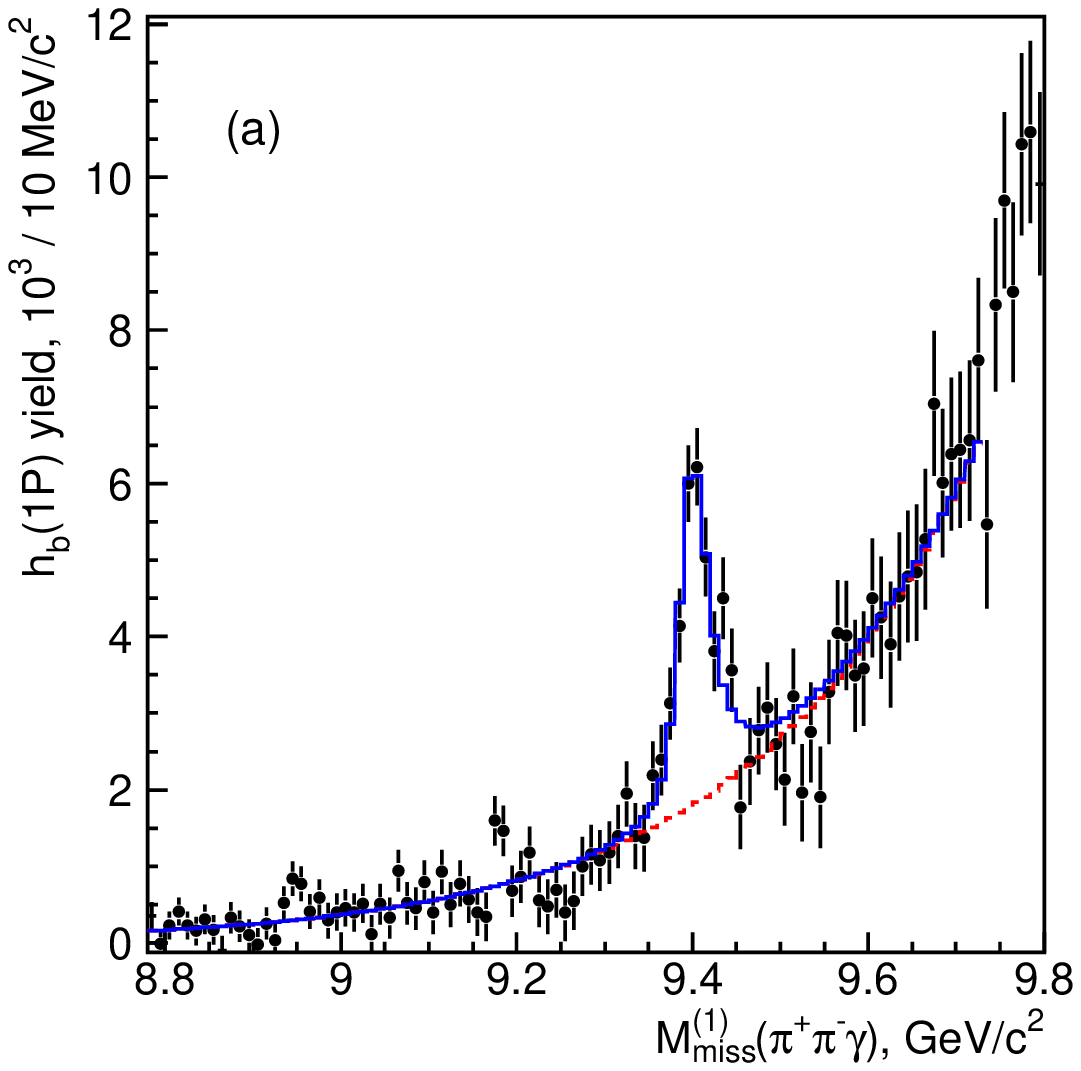}\hfill
\includegraphics[width=0.33\linewidth]{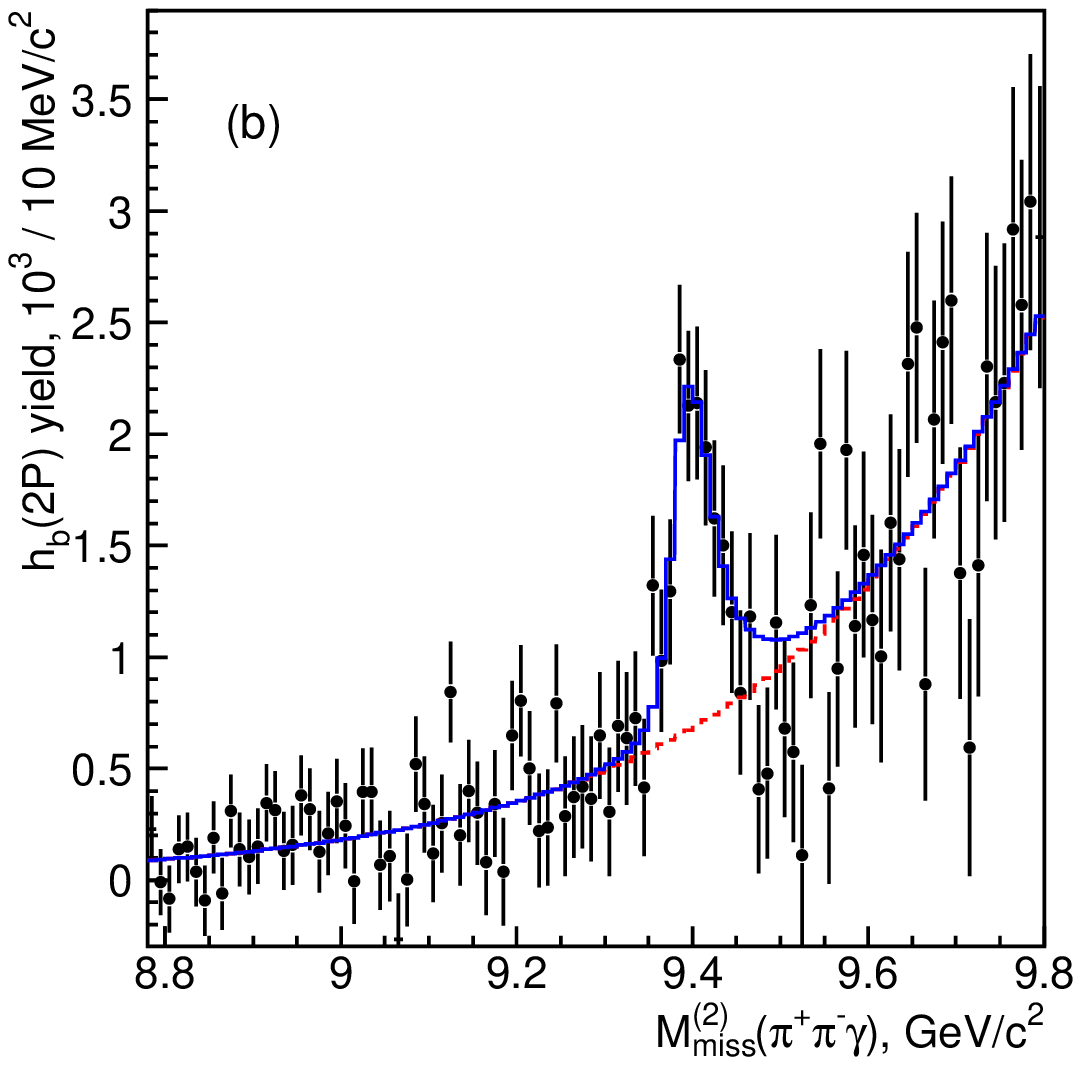}\hfill
\includegraphics[width=0.33\linewidth]{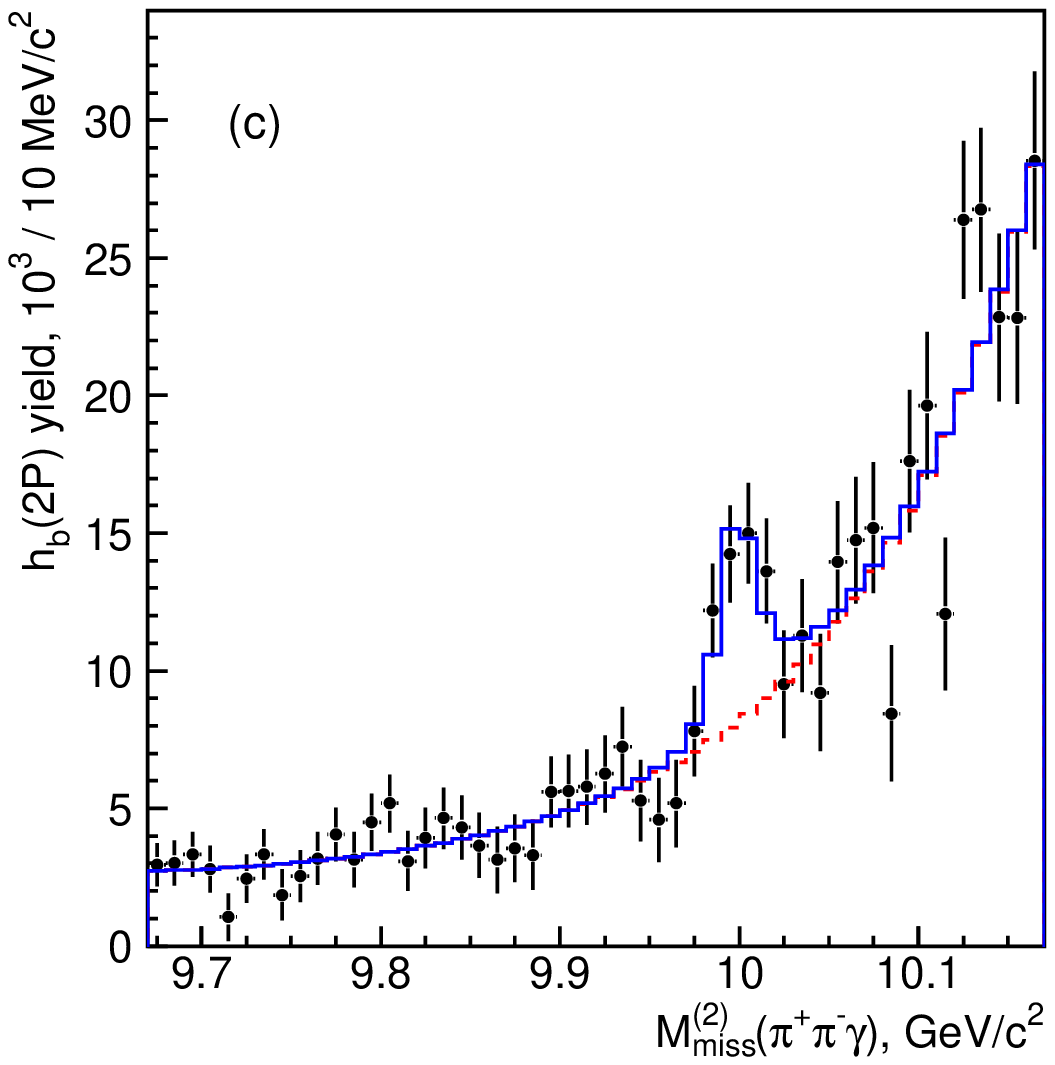}
\caption{ The $\hb$ (a) and $\hbp$ (b), (c) yields as a function of
  the $\pp\ga$ missing mass. }
\label{fig:et_belle}
\end{figure}

Large samples of the $\hb$ and $\hbp$ enable the study of the $\et$
and $\ett$ states, since the electric-dipole transitions
$\hbn\to\etm\ga$ are expected to be prominent~\cite{god_ros}.
To reconstruct these transitions Belle measured the $\hbn$ yield as a
function of the $\pp\ga$ missing mass~\cite{et_belle}. The radiative
transitions from the $\hb$ and $\hbp$ to the $\et$ are observed with
significances of $15\,\sigma$ and $9\,\sigma$, respectively [see
  Fig.~\ref{fig:et_belle}~(a) and (b)].
From simultaneous fit the mass and width of the $\et$ are measured to
be $m_{\et}=(9402.4\pm1.5\pm1.8)\,\mevm$ and
$\Gamma_{\et}=(10.8\,^{+4.0}_{-3.7}\,^{+4.5}_{-2.0})\,\mev$.
The $\Gamma_{\et}$ is a first measurement; the $m_{\et}$ measurement
is more precise than the current world average and is
$(11.4\pm3.6)\,\mevm$ above the central value~\cite{PDG}.
The measured hyperfine splitting, $\Delta M_{\rm
  HF}(1S)=(57.9\pm2.3)\,\mevm$, is in agreement with perturbative
NRQCD $(41\pm14)\,\mevm$~\cite{qwg_reports,et_pqcd} and Lattice
$(60\pm8)\,\mevm$~\cite{et_lattice} calculations.

Belle found first evidence for the $\ett$ using radiative transition
from the $\hbp$ [see Fig.~\ref{fig:et_belle}~(c)]. The $\ett$
significance is $4.4\,\sigma$ including systematic uncertainties and
``look elsewhere'' effect. The mass of the $\ett$ is measured to be
$m_{\ett}=(9999.0\pm3.5\,^{+2.8}_{-1.9})\,\mevm$, the hyperfine
splitting is $\Delta M_{\rm HF}(2S)=(24.3^{+4.0}_{-4.5})\,\mevm$. For
the ratio of hyperfine splittings many theoretical uncertainties
cancel. Belle measurement $\Delta M_{\rm HF}(2S)/\Delta M_{\rm
  HF}(1S)=0.420^{+0.071}_{-0.079}$ is in agreement with NRQCD, Lattice
and model-independent
calculations~\cite{qwg_reports,et_pqcd,et_lattice,et_lattice2,et_model_indep}.

Belle measured also branching fractions 
$\br[\hb\to\et\gamma]=(49.2\pm5.7\,^{+5.6}_{-3.3})\%$,\\
$\br[\hbp\to\et\gamma]=(22.3\pm3.8\,^{+3.1}_{-3.3})\%$ and
$\br[\hbp\to\ett\ga]=(47.5\pm10.5\,^{+6.8}_{-7.7})\%$.
These branching fractions are somewhat higher than the quark model
predictions~\cite{god_ros}.

There is another claim of the $\ett$ signal by the group of K.~Seth
from Northwestern University, that used CLEO data~\cite{et_kamal}. The
$\Ut\to\ett\ga$ production channel is considered and the $\ett$ is
reconstructed in 26 exclusive channels with up to 10 charged tracks in
the final state. This analysis requires reconstruction of low energy
photon, $E_{\ga}\approx\Delta M_{\rm HF}(2S)$. Claimed significance of
the $\ett$ is $4.6\,\sigma$. The group measured $\Delta M_{\rm
  HF}(2S)=(48.7\pm2.7)\,\mevm$. This value is two times higher then
the Belle measurement and inconsistent with Belle at the $5\,\sigma$
level. We would like to stress, that for the Belle value of the
$\Delta M_{\rm HF}(2S)$ the background in the $\Ut\to\ett\ga$ channel
is very high, thus this reaction has no sensitivity to low values of
the $\Delta M_{\rm HF}(2S)$ and the two reported signals can not have
the same origin. We point out that in~\cite{et_kamal} the background
is assumed to depend exponentially on the $\ga$ energy, while it is
known that the final state radiation contributes power law tail, thus
the background model is incomplete and the claimed significance is
overestimated. In addition, counting the event yields in the figure
from~\cite{et_kamal} one can conclude that
$N[\Ut\to\ett\ga]\approx0.2\,N[\Ut\to\chi_{b1}(1P)\ga]$. Analogous
process in charmonium sector $\psp\to\eta_c(2S)\ga$ was recently
observed by BESIII~\cite{besiii_ettc} and the relevant ratio
$\br[\psp\to\eta_c(2S)\ga]\approx0.005\,\br[\psp\to\chi_{c1}\ga]$ is
by a factor 30 lower. We conclude that the signal claimed
in~\cite{et_kamal} is unlikely to originate from the $\ett$.

\subsection{Evidence for $\psi_2(1D)$}

The $D$-wave charmonium levels are expected to be situated between the
$D\bar{D}$ and $D\bar{D}^*$ thresholds~\cite{d-wave-charmonium}. Among
them, the $\eta_{c2}$ with $J^{PC}=2^{-+}$ and $\psi_2$ with
$J^{PC}=2^{--}$ have unnatural spin-parities and can not decay to
$D\bar{D}$. Thus they remain the only undiscovered narrow charmonia.

\begin{figure}[tbhp]
\center
\includegraphics[width=0.5\linewidth]{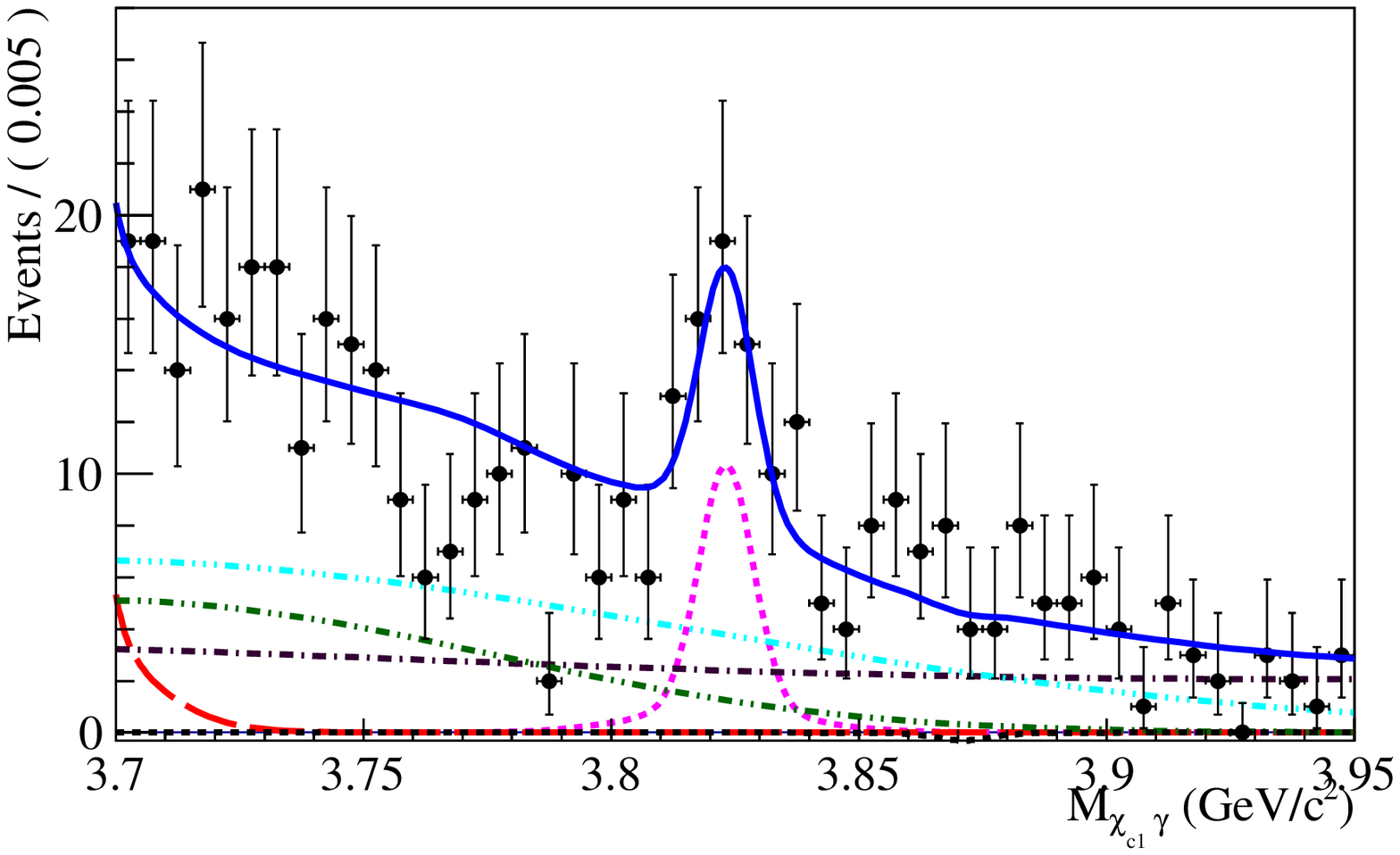}
\caption{ The $M(\chi_{c1}\ga)$ spectrum for the $B^+\to
  K^+\chi_{c1}\ga$ candidates. }
\label{fig:psi_2_1d}
\end{figure}

Belle reported preliminary results on the resonant structure of the
$B^+\to K^+\chi_{c1}\ga$ decays, with $\chi_{c1}$ reconstructed in the
$\jp\ga$ mode. In the channel $\chi_{c1}\ga$ Belle finds the first
evidence for the $\psi_2(1D)$ charmonium state (see
Fig.~\ref{fig:psi_2_1d}),
with the mass of $M=(3823.5\pm2.8)\,\mevm$ and the significance of
$4.2\,\sigma$ including systematic uncertainty. Measured width is
consistent with zero, $\Gamma=(4\pm6)\,\mev$; it is likely that the
width is very small, since the state is observed in the radiative
decay and the typical charmonium radiative decay widths are at the
$O(100)\,\kev$ level. The odd $C$-parity (fixed by decay products)
allows to discriminate between the $\eta_{c2}$ and $\psi_2$
hypotheses. No signal is found in the $\chi_{c2}\ga$ channel, in
agreement with expectations for the $\psi_2$~\cite{d-wave-charmonium}.

Belle measures $\br[B^+\to K^+\psi_2]\times\br[\psi_2\to\chi_{c1}\ga]=
(9.7{^{+2.8}_{-2.5}}{^{+1.1}_{-1.0}})\times10^{-6}$. Given expected
$\br[\psi_2\to\chi_{c1}\ga]\sim\frac{2}{3}$~\cite{d-wave-charmonium},
the $\br[B^+\to K^+\psi_2]$ is a factor 50 smaller than corresponding
branching fractions for the $\jp$, $\psp$ and $\chi_{c1}$ due to the
factorization suppression~\cite{fact-supp}.

\section{Quarkonium(-like) states above open flavor thresholds}

Many recently observed states above open flavor thresholds exhibit
anomalously large rates of transitions to lower quarkonia with
emission of light hadrons.

There are five such states in charmonium sector: the
$Y(3915)\to\jp\omega$ observed in $B$ meson decays and in $\ga\ga$
fusion~\cite{y3915_belle_babar}, and four states observed in the
initial state radiation (ISR) process:
$Y(4008,4260)\to\jp\pp$~\cite{y4260_babar_belle} and
$Y(4360,4660)\to\psp\pp$~\cite{y4360_babar_belle}. 

Other charmonium(-like) states [$\psi(3770)$, $\psi(4040)$,
  $\psi(4160)$, $\psi(4415)$, $X(3940)$, $X(4160)$] decay to open
flavor channels. The $\psi$ states are successfully interpreted as
charmonium levels, while the $X(3940)$ and $X(4160)$ states, observed
in double-charmonium production process~\cite{double_ccbar_belle},
have masses about $100\,\mevm$ away from likely $c\bar{c}$
assignments. Hadronic transitions from these states to lower charmonia
were not known [except for $\psi(3770)$].

\subsection{ Observation of $\psi(4040)$ and $\psi(4160)$ transitions
  to $\jp\eta$}

\begin{figure}[tbhp]
\center
\includegraphics[width=0.33\linewidth,angle=-90]{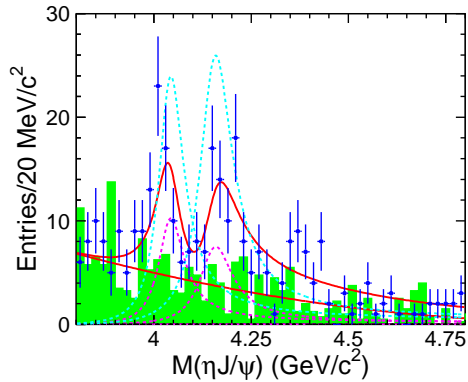}
\caption{The $\eta\jp$ invariant mass distribution and the fit
  results. The points with error bars show the data while the shaded
  histogram is the normalized $\eta$ and $\jp$ background from the
  sidebands. The curves show the best fit on signal candidate events
  and sideband events simultaneously and the contribution from each
  Breit-Wigner component.}
\label{fig:isr_jp_eta}
\end{figure}

Recently Belle observed transitions from $\psi(4040)$ and $\psi(4160)$
to $\jp\eta$ (see Fig.~\ref{fig:isr_jp_eta}) using the ISR scan of the
$\ee\to\jp\eta$ cross-section~\cite{jp_eta_belle}.
The partial widths of these transitions are $\Gamma\sim1\,\mev$, which
is anomalously large. For the first time the $\psi$ states that are
considered to be ``conventional charmonia'' show anomalous
properties.

Similar phenomenon was found also in bottomonium sector.

\subsection{Anomalous hadronic transitions from $\Uf$}

In 2008 Belle observed anomalously large rates of the $\Uf\to\Un\pp$
($n=1,2,3$) transitions with partial widths of
$\Gamma=300-400\,\kev$~\cite{uf_pp_belle}. This can be compared with
$\Gamma=1-6\,\kev$ for transitions from lower $\Un$ states
($n=2,3,4$)~\cite{PDG}. Recently Belle found more hadronic transitions
from the $\Uf$ state.

\subsubsection{Observation of $\Uf\to\U(1D)\pp$}

Already in the $\hbn$ analysis Belle found signal of the inclusively
reconstructed decay $\Uf\to\U(1D)\pp$ with the marginal significance
of $2.6\,\sigma$~\cite{hb_belle}. The ratio of the $\U(1D)$ and $\Ut$
yields is found to be $\sim1/7$, thus partial decay width
$\Gamma[\Uf\to\U(1D)\pp]$ is at the level of $60\,\kev$, which is
anomalously large.

\begin{figure}[tbhp]
\includegraphics[width=0.42\linewidth]{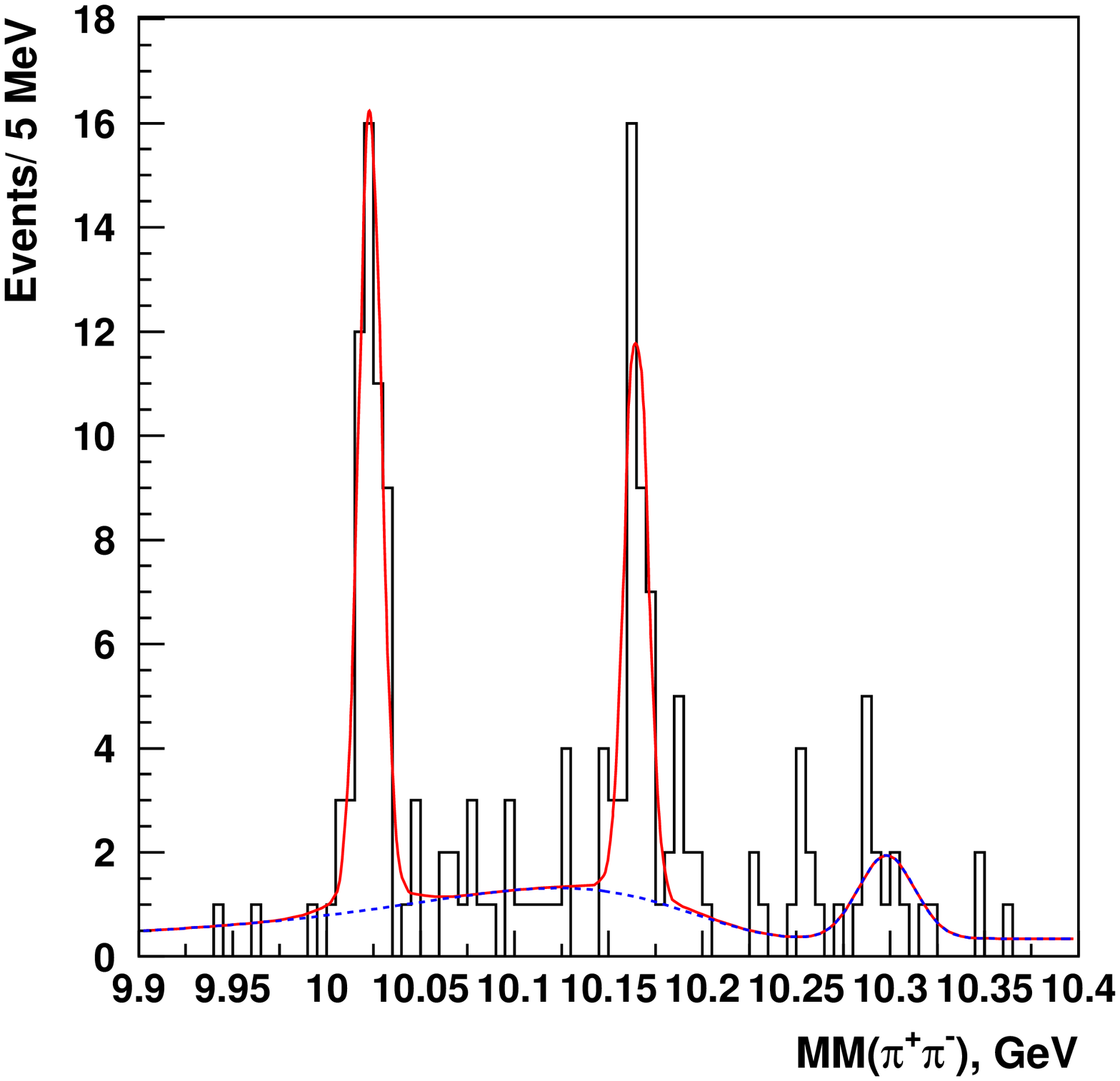}\hfill
\includegraphics[width=0.42\linewidth]{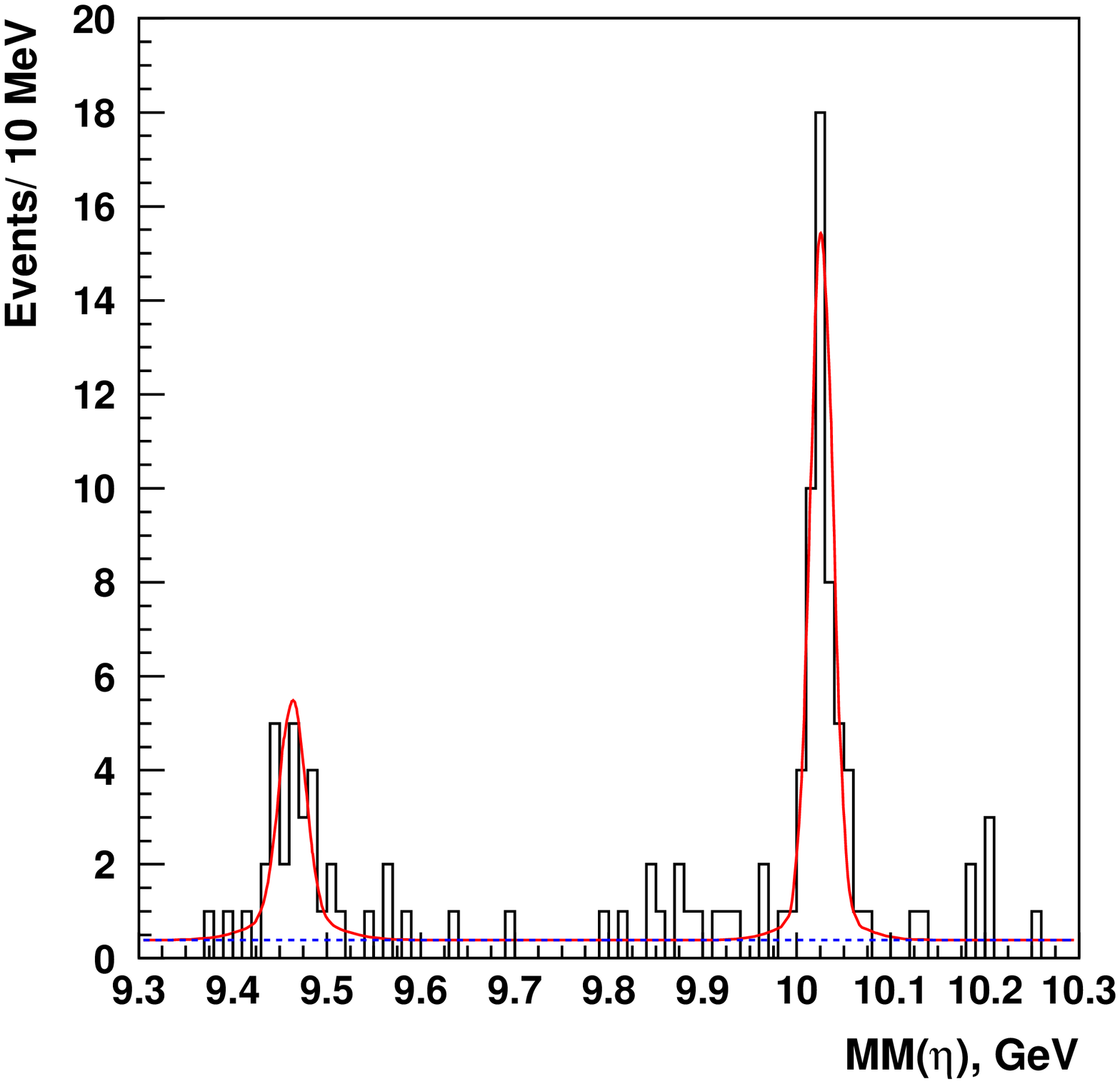}
\caption{ The missing mass spectra of $\pp$ (left) and $\eta$ (right)
  for the exclusively reconstructed hadronic transitions from the
  $\Uf$. }
\label{fig:uf_hadr}
\end{figure}

Exclusive reconstruction of the decay chain $\Uf\to\U(1D)\pp$,
$\U(1D)\to\chi_{bJ}(1P)\ga$, \\ 
$\chi_{bJ}(1P)\to\Uo\ga$, $\Uo\to\uu$ allowed to observe this
transition with the significance of $9\,\sigma$, according to
preliminary Belle results (see Fig.~\ref{fig:uf_hadr} left panel).
Belle measures product of branching fractions:
$\br[\Uf\to\U(1D)\pp]\times(\sum\limits_{J=1,2}
\br[\U(1D)\to\chi_{bJ}\ga]\times\br[\chi_{bJ}\to\Uo\ga])=
(2.5\pm0.5\pm0.5)\times10^{-5}$. This is 10 times higher than a
similar product for the $\Uth\to\U(1D)\ga\ga$ production channel that
was observed by CLEO~\cite{y1d_cleo}.

\subsubsection{Observation of $\Uf\to\U(1S,2S)\eta$}

In the multipole expansion the $\eta$ transitions are dominated by the
emission of $E1M2$ gluons, while the $\pp$ transitions are dominated
by $E1E1$ gluons, thus the ratio $R_{n\to
  m}=\frac{\U(nS)\to\U(mS)\eta}{\U(nS)\to\U(mS)\pp}$ is expected to be
small~\cite{multip_expans}. This was indeed observed for low $\U$
states, {\it e.g.} Belle recently measured $R_{2\to
  1}=(1.99\pm0.14^{+0.12}_{-0.08})\times10^{-3}$~\cite{y2s_eta_belle}.
In 2008 BaBar observed the $\U(4S)\to\eta\U(1S)$ transition and
measured $R_{4\to 1}=2.41\pm0.40\pm0.21$~\cite{y4s_eta_babar}. This
unexpectedly high value pointed out to the breakdown of the multipole
expansion approach.

Recently Belle reported preliminary results on the observation of the
$\Uf\to\U(1S,2S)\eta$ transitions (see Fig.~\ref{fig:uf_hadr} right
panel). Belle measures the branching fractions
$\br[\Uf\to\Uo\eta]=(0.73\pm0.16\pm0.08)\times10^{-3}$,
$\br[\Uf\to\Uo\eta]=(3.8\pm0.4\pm0.5)\times10^{-3}$, which translate
to $40\,\kev$ and $200\,\kev$ partial widths, respectively; such
widths are anomalously high. For the relative rates Belle found
$R_{5\to 1}=0.16\pm0.04\pm0.02$ and $R_{5\to
  2}=0.48\pm0.05\pm0.09$. Thus, there is no strong suppression of
$\eta$ transitions from the $\Uf$ relative to $\pp$ transitions.

\subsection{Interpretation}

It is proposed that the anomalously large rates of the hadronic
transitions from the quarkonia above open flavor thresholds are due to
the contribution of the hadron
loops~\cite{loops_simonov,loops_meng}. The phenomenon can be
considered either as a rescattering of the $D\bar{D}$ or $B\bar{B}$
mesons, or as a contribution of the four-quark molecular component to
the quarkonium wave-function. Unsuppressed $\eta$ transitions could
have similar explanation~\cite{eta_voloshin}.

Despite striking similarity between the observations in the charmonium
and bottomonium sectors, there is also some difference. In charmonium,
each of the $Y(3915)$, $Y(4008)$, $\psi(4040)$, $\psi(4160)$,
$Y(4260)$, $Y(4360)$ and $Y(4660)$ states decays to only one
particular channel [$\jp\omega$, $\jp\eta$, $\jp\pp$ or $\psp\pp$]. In
bottomonium, we know only one state with anomalous properties, the
$\Uf$, that decays to many different channels [$\Un\pp$, $\hbm\pp$,
  $\U(1D)\pp$, $\Un\eta$] with similar probabilities for each
channel. There is no general model giving explanation to this
difference between charmonium and bottomonium. To explain affinity of
the charmonium-like states to some particular channels, the notion of
``hadrocharmonium'' was proposed~\cite{hadroquarkonium_voloshin}. It
is a heavy quarkonium embedded into a cloud of light hadron(s), thus
the fall-apart decay could be dominant. Hadrocharmonium could also
provide an explanation for the charged charmonium-like states observed
by Belle~\cite{zc_belle}.

\section{Summary}

There are many new results from Belle on the heavy quarkonium.  The
number of spin-singlet bottomonium states has increased from one to
four over the last two years, including more precise measurement of
the $\et$ mass which appeared to be $11\,\mevm$ away from the PDG2012
average. There is an evidence of one of the two still missing narrow
charmonium states expected in the region between the $D\bar{D}$ and
$D\bar{D}^*$ thresholds. 

Observation and detailed studies of the {\it charged} bottomonium-like
states $\zbo$ and\\ $\zbt$ (discussed in a separate talk) open reach
phenomenological field to study exotic states near open flavor
thresholds.

Belle observed new decay channels of the charmonium- and
bottomonium-like states above open flavor thresholds. General feature
of these states is the large rates of transitions to lower quarkonia
with the emission of light hadrons. Hadron loops are important for
understanding of their properties, however, there is no general
theoretical model for these highly excited states yet.

We conclude that properties of low excitations are in agreement with
the Lattice QCD and effective theories calculations, while high
excitations show some unexpected properties, which are still not well
understood.  Interestingly, similar phenomena near and above open
flavor thresholds are found in bottomonium and charmonium sectors.

\end{document}